\documentclass[aps,pra,reprint,showpacs]{revtex4-1}
\pdfoutput=1

\usepackage[pdftex]{graphicx}
\usepackage{times}
\usepackage{amsmath}
\usepackage{amsfonts}
\usepackage{mathrsfs}
\usepackage{bm}
\usepackage{hyperref}

\begin{document}

\title{Comment on ``Influence of induced interactions on superfluid properties of quasi-two-dimensional dilute Fermi gases with spin-orbit coupling"}

\author{Juhee Lee}
\affiliation{Department of Physics and Photon Science, School of Physics and Chemistry, Gwangju Institute of Science and Technology, Gwangju 61005, Korea}
\author{Dong-Hee Kim}
\email{dongheekim@gist.ac.kr}
\affiliation{Department of Physics and Photon Science, School of Physics and Chemistry, Gwangju Institute of Science and Technology, Gwangju 61005, Korea}

\begin{abstract}
In an article in 2013, Caldas {\it et al.} [\href{http://dx.doi.org/10.1103/PhysRevA.88.023615}{Phys. Rev. A 88, 023615 (2013)}] derived analytical expressions of the induced interaction within the scheme of Gorkov and Melik-Barkhudrov in quasi-two-dimensional Fermi gases with Rashba spin-orbit coupling (SOC). They claimed that the induced interaction is exactly the same as the one for the case without SOC when the SOC is weak, and in the region of strong SOC, it starts from a reduced value and then recovers the value for the zero SOC in the limit of large SOC. We point out that their calculations contain the critical errors and inconsistencies that significantly affect the basis of these claims.     
\end{abstract}


\maketitle

Caldas {\it et al.}~\cite{Caldas2013} calculated the induced interaction for attractively interacting Fermi gases with Rashba spin-orbit coupling (SOC) in two dimensions. They provided the first calculation considering the Gorkov--Melik-Barkhudarov (GMB) correction to the superfluid properties in presence of the Rashba SOC. The accurate estimation of superfluid transition temperature is of clear importance in ultracold gas systems where the realization of the SOC becomes possible in a controllable environment (for instance, see \cite{Huang2016}).   
In \cite{Caldas2013}, they obtained the analytical expressions of the induced interaction in the weak and strong SOC regimes, claiming that (i) the magnitude of the induced interaction is exactly the same as the value for the case without SOC in the weak coupling regime, and (ii) in the strong SOC regime, the magnitude starts from a reduced value but recovers the zero-SOC case in the limit of large SOC. Unfortunately, these claims are flawed because of the critical errors and physical inconsistencies found in their calculations.

The claim (i) for the weak SOC regime ($\lambda \ll v_F$) was deduced from the result of Eq. (3.18) of \cite{Caldas2013}, 
\begin{eqnarray*}
F(\alpha) &=& \frac{1}{2\pi} \int_0^{2\pi} d\theta \left[ 1 + \frac{4}{1+\cos \theta} \alpha^2 \right] \\
&=& 1 + 4\alpha^2 \frac{1}{2\pi} \int_0^{2\pi} d\theta  \frac{1}{1+\cos \theta} \\
&=& 1 + \frac{4\alpha^2}{2\pi} \tan (\theta /2 ) \big\vert_0^{2\pi} = 1.
\end{eqnarray*} 
This calculation is incorrect since $\int_0^{2\pi} d\theta \frac{1}{1+\cos \theta}$ in fact diverges. Therefore, the induced interaction $\bar{U}_{\mathrm{int}}(\lambda) = g^2 N(0) F(\alpha)$ in Eq. (3.17) diverges to infinity for any finite SOC strength $\alpha$, and the claim (i) loses its ground.

The claim (ii) for the strong SOC regime ($\lambda \gg v_F$) stemmed from the derivation of the polarization function $\chi(q,\lambda)$ given in Eq. (3.20) of  \cite{Caldas2013} as
\begin{equation*}
\chi(q,\lambda)=-N(0)\left[ \sqrt{1+\left(\frac{2k_F}{b}\right)^2} + \frac{q^2}{2b^2} \ln \left( \frac{q^2 - 2bk_F^+}{q^2 + 2bk_F^-} \right) \right]
\end{equation*}  
where $k_F^+$ and $k_F^-$ are the magnitudes of the Fermi momenta of the ($+$) and ($-$) helicity branches with energy dispersion $\xi_{\mathbf{k},\pm} = \xi_\mathbf{k} \pm \lambda k$. However, at a fixed density of particles as assumed in  \cite{Caldas2013}, in the BEC regime with the strong SOC ($\mu<0$ at large $\lambda$),  $k_F^+$ does not exist since the Fermi sea forms only with particles in the ($-$) helicity branch (for instance, see \cite{Shi2016} and Fig.~\ref{fig1} below). 

\begin{figure}[b]
\includegraphics[width=0.38\textwidth]{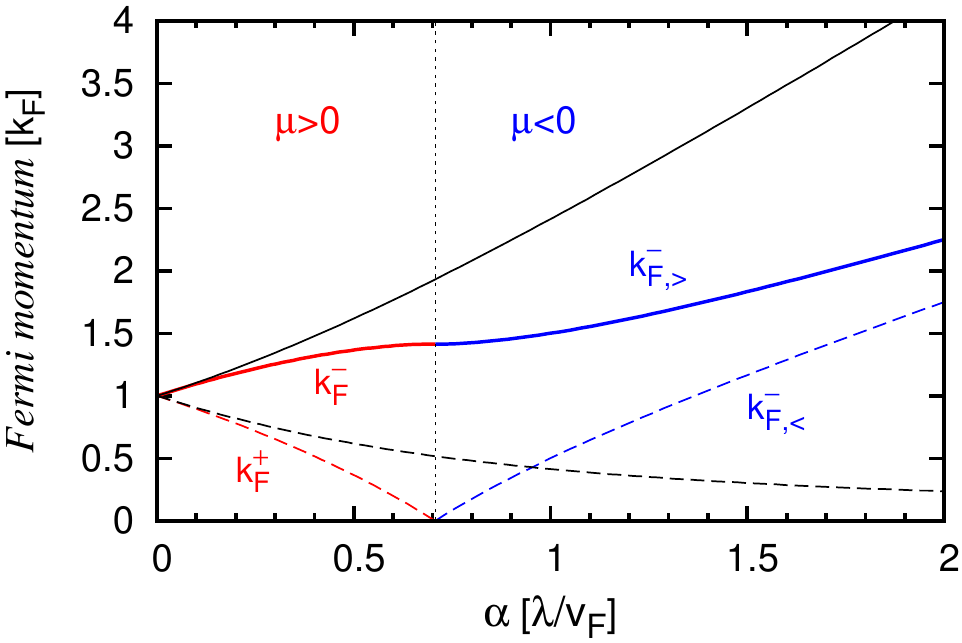}
\caption{(Color online) Fermi momentum of a two-dimensional Fermi gas with Rashba SOC. The particle density is fixed as $n=k_F^2/2\pi$. The sign change of chemical potential $\mu$ is marked by the vertical dotted line at $\lambda/v_F=1/\sqrt{2}$. At $\mu<0$, the Fermi sea is only from the (-) helicity branch and doughnut-like shaped with inner radius $k_{F,<}^-$ and outer radius $k_{F,>}^-$. The black solid and dotted lines of $k_F^\pm / k_F = \pm \alpha + \sqrt{\alpha^2 + 2m\mu/k_F^2}$ [Eq. (3.10)] with the use of $k_F=\sqrt{2m\mu}$ shown in Fig. 4 of \cite{Caldas2013} are given for direct comparison.
\label{fig1}}
\end{figure}

This issue is connected to the inconsistent use of $k_F$ in the earlier part of the article. In the end of Section II of \cite{Caldas2013}, they defined $k_F$ as $n=k_F^2/2\pi$ with the particle density $n$ which they clarified is fixed throughout their calculations. However, in Fig. 4 of \cite{Caldas2013} plotting $k_F^\pm/k_F$, they apparently used $k_F=\sqrt{2m\mu}$, which is also found in the earlier part of the article, but it is not equivalent to $k_F=\sqrt{2\pi n}$ for any finite SOC. At a fixed $n$, the number equation [Eq. (2.11)] can be solved for chemical potential $\mu$ in the noninteracting and zero-temperature limit. For the BCS (weak SOC) regime ($\mu>0$), it leads to $\mu=\epsilon_F - m\lambda^2$; for the BEC (strong SOC) regime ($\mu<0$), it gives $\mu=-\frac{1}{2}m\lambda^2+\frac{\epsilon_F^2}{2m\lambda^2}$, where $\epsilon_F=k_F^2/2m$, verifying that $k_F=\sqrt{2\pi n}\neq \sqrt{2m\mu}$. We provide the corrected Fermi momenta for non-zero SOC in Fig.~\ref{fig1}. 

Another inconsistency is found in the $\mu$-calculation shown in Fig. 7 of \cite{Caldas2013}. They compared their calculation of $\mu$ with the previous mean-field result \cite{Chen2012}, which shows good agreement with the $x$-axis value $\sim 1.4$ at which $\mu$ changes its sign. However, in fact the units are different between \cite{Caldas2013} and \cite{Chen2012}: in \cite{Caldas2013}, it was $\lambda/v_F$ in the $x$-axis, while it was $\lambda k_F /\epsilon_F$ in Fig. 1 of \cite{Chen2012}, which indicates a factor of two difference between the two. Therefore, in order to be consistent with the previous work~\cite{Chen2012}, $\mu=0$ was supposed to be found at $\lambda/v_F\sim 0.7$. The value $0.7$ agrees with our estimation of $\lambda/v_F=1/\sqrt{2}$ shown in Fig.~\ref{fig1} for the noninteracting case.   

The errors and inconsistencies shown above suggest a possibility of the errors being propagated from the very early stage of evaluating the polarization function $\chi_{ph}$, i.e. Eqs. (3.3) and (3.9). While reproducing Eq. (3.9) is not straightforward, the discrepancy between Eqs. (3.3) and (3.9) is indeed identified in the limit of $\mathbf{q}=0$ at finite $\lambda$ and $\mu>0$. For $\mathbf{q}\to 0$, Eq. (3.3) can be evaluated as
\begin{equation*}
\chi_{ph}(q\to 0,\lambda) = \int \frac{d^2 \mathbf{k}}{(2\pi)^2} \frac{f_{\mathbf{k}}^- - f_{\mathbf{k}}^+}{\xi_\mathbf{k}^+ - \xi_\mathbf{k}^-} = \frac{1}{4\pi\lambda} \left( k_F^- - k_F^+ \right),
\end{equation*}
while Eq. (3.9) gives a very different result as
\begin{equation*}
-\frac{m}{2\pi} \int_0^\infty dk k \frac{f_{\mathbf{k}}^- + f_{\mathbf{k}}^+}{2\lambda m k} = -\frac{1}{4\pi\lambda} \left( k_F^- + k_F^+ \right).
\end{equation*}
This implies that intermediate steps of deriving (3.9) from (3.3) that were actually not given in \cite{Caldas2013} may contain critical errors. Furthermore, the expansion for $\lambda \ll v_F$ is not correct, either. In Eq (3.15), 
$\chi(q,\lambda) = -N(0) \left[ 1 + 8 \frac{m^2\lambda^2}{q^2} \right]$,
the evaluation in the limit of $\mathbf{q}\to 0$ diverges, which is not consistent with its original formula in Eq. (3.9) that is anyway finite at $\mathbf{q}=0$. 

The discrepancy between Eqs. (3.3) and (3.9) can be also seen in an alternative way. Let us focus only on the factor in the first term with $f_k^-$ of the integrand in Eq. (3.9). The corresponding angular integration in Eq. (3.3) to derive this factor can be explicitly written as 
\begin{widetext}
\begin{equation*}
\int_0^{2\pi} \frac{d\phi}{(2\pi)^2} \frac{1}{\xi_\mathbf{k}^+ - \xi_{\mathbf{k}+\mathbf{q}}^-} =
\frac{1}{(2\pi)^2}\int_0^{2\pi} d\phi \frac{1}{\left[ \frac{k^2}{2m} + \lambda k \right] - \left[ \frac{k^2+q^2+2kq\cos\phi}{2m} - \lambda \sqrt{k^2+q^2+2kq\cos\phi} \right]},
\end{equation*} 
\end{widetext} 
where $\phi$ is an angle between $\mathbf{k}$ and $\mathbf{q}$. Since this integration is well-defined for all $k \in [0,\infty)$, one can simply check the consistency with Eq. (3.9) in the limit of $k\to 0$ for finite $q$. In this limit, the angle dependence is irrelevant, and thus the integration becomes $\frac{m}{\pi} \frac{1}{-q^2+2m\lambda q}$. 
In contrast, by directly evaluating the limit of $k \to 0$ in the corresponding factor in Eq. (3.9), one finds $-\frac{m}{\pi q^2}$ where $\lambda$ does not appear. This disappearance of $\lambda$ cannot be explained since there is no source of the cancellation of $\lambda$ in this part of the evaluation in Eq. (3.3). 

In addition, Eq. (3.3) may have a typo. The Matsubara frequency summation giving the second line of Eq. (3.3) is not consistent with the known form of the polarization function evaluated in the previous studies of the induced interaction correction~\cite{Heiselberg2000,Petrov2003,Baranov2008,Kim2009,Martikainen2009,Yu2010}. It should be corrected as 
$\frac{d^2\mathbf{k}}{(2\pi)^2} \frac{f_\mathbf{k}^- - f_{\mathbf{k}+\mathbf{q}}^+}{i\Omega_l + \xi_\mathbf{k}^- - \xi_{\mathbf{k}+\mathbf{q}}^+}$.
However, this typo correction cannot resolve the inconsistencies discussed above, and again it is very likely that Eq. (3.9) contains nontrivial errors from the intermediate steps of the angular integration which unfortunately were not shown in \cite{Caldas2013}.

In conclusion, we have pointed out that the calculations of the induced interaction in \cite{Caldas2013} contain derivation errors and inconsistencies that critically affect the main claims of the article.


\begin{thebibliography}{}

\bibitem{Caldas2013}
H. Caldas, R. L. S. Farias, and M. Continentino, \href{http://dx.doi.org/10.1103/PhysRevA.88.023615}{Phys. Rev. A {\bf 88}, 023615 (2013)}.

\bibitem{Huang2016}
L. Huang, Z. Meng, P. Wang, P. Peng, S.-L. Zhang, L. Chen, D. Li, Q. Zhou, and J. Zhang,
\href{http://dx.doi.org/10.1038/nphys3672}{Nat. Phys. {\bf 12}, 540 (2016)}.

\bibitem{Shi2016}
H. Shi, P. Rosenberg, S. Chiesa, and S. Zhang,
\href{http://dx.doi.org/10.1103/PhysRevLett.117.040401}{Phys. Rev. Lett. {\bf 117}, 040401 (2016)}.

\bibitem{Chen2012}
G. Chen, M. Gong, and C. Zhang, \href{http://dx.doi.org/10.1103/PhysRevA.85.013601}{Phys. Rev. A {\bf 85}, 013601 (2012)}.

\bibitem{Heiselberg2000}
H. Heiselberg, C. J. Pethick, H. Smith, and L. Viverit, 
\href{http://dx.doi.org/10.1103/PhysRevLett.85.2418}{Phys. Rev. Lett. {\bf 85}, 2418 (2000)}.

\bibitem{Petrov2003}
D. S. Petrov, M. A. Baranov, and G. V. Shlyapnikov, 
\href{http://dx.doi.org/10.1103/PhysRevA.67.031601}{Phys. Rev. A {\bf 67}, 031601(R) (2003)}.

\bibitem{Baranov2008}
M. A. Baranov, C. Lobo, and G. V. Shlyapnikov, 
\href{http://dx.doi.org/10.1103/PhysRevA.78.033620}{Phys. Rev. A {\bf 78}, 033620 (2008)}.

\bibitem{Kim2009}
D.-H. Kim, P. T\"orm\"a, and J.-P. Martikainen, 
\href{http://dx.doi.org/10.1103/PhysRevLett.102.245301}{Phys. Rev. Lett. {\bf 102}, 245301 (2009)}. 

\bibitem{Martikainen2009}
J.-P. Martikainen, J. J. Kinnunen, P. T\"orm\"a, and C. J. Pethick, 
\href{http://dx.doi.org/10.1103/PhysRevLett.103.260403}{Phys. Rev. Lett. {\bf 103}, 260403 (2009)}.

\bibitem{Yu2010}
Z.-Q. Yu and L. Yin, 
\href{http://dx.doi.org/10.1103/PhysRevA.82.013605}{Phys. Rev. A {\bf 82}, 013605 (2010)}.

\end{thebibliography}
\end{document}